\documentclass[aps,prb,superscriptaddress,twocolumn,showpacs]{revtex4}
\usepackage{color}
\usepackage{amsmath}
\usepackage{graphicx}
\usepackage{amsfonts}
\usepackage{amssymb}
\usepackage{subfigure}
\usepackage{float}
\usepackage{amssymb}

\begin{document}

\title{Aharonov-Casher effect in quantum ring ensembles}
\author{Fateme K. Joibari}
\affiliation{Kavli Institute of NanoScience, Delft University of Technology, Delft, The
Netherlands}
\author{Ya. M. Blanter}
\affiliation{Kavli Institute of NanoScience, Delft University of Technology, Delft, The
Netherlands}
\author{ Gerrit E. W. Bauer}
\affiliation{Kavli Institute of NanoScience, Delft University of Technology, Delft, The
Netherlands}
\affiliation{Institute for Materials Research and WPI-AIMR, Tohoku University, Sendai,
Japan}
\date{\today}

\begin{abstract}
We study the transport of electrons through a single-mode quantum ring with electric field induced Rashba spin-orbit interaction 
that is subjected to an in-plane magnetic field and weakly coupled to electron reservoirs. Modelling a ring array by ensemble averaging
 over a Gaussian distribution of energy level positions, we predict slow conductance oscillations as a function of the Rashba interaction 
and electron density due to spin-orbit interaction-induced beating of the spacings between the levels crossed by the Fermi energy. Our results agree with experiments by Nitta c.s.,
 thereby providing an interpretation that differs from the ordinary Aharonov-Casher effect in a single ring.
\end{abstract}
\pacs{73.23.-b, 73.63.-b, 71.70.Ej}
\maketitle

The Aharonov-Casher (AC) effect\cite{AC} is an analog of the Aharonov-Bohm
(AB) effect, but caused by the spin-orbit interaction (SOI) rather than an
external magnetic field. Originally, Aharonov and Casher predicted in 1984
that a spin accumulates a phase when the electric charge is circling in an
external electric field.\cite{AC} This situation is similar to a single-mode
ballistic ring with the Rashba spin-orbit interaction. Quantum rings in
high-mobility semiconductor material have therefore attracted extensive
attention, both experimentally and theoretically, as model devices to
investigate fundamental quantum mechanical phenomena.

In the ordinary AC effect, the electrons injected into a quantum ring with
SOI acquire spin phases when traversing the two arms due to precession in
the effective spin-orbit magnetic field. Interference of the spinor wave
functions at the exit point of the ring then leads to an oscillatory
conductance as a function of the spin-orbit coupling constant that in Rashba
systems can be tuned by an external gate voltage. The theory of AC
conductance oscillations\cite{Diego} in a single-mode quantum ring
symmetrically coupled to two leads is in a good agreement with experimental
observations.\cite{Bergsten} More recently, the zero magnetic field
conductance behavior as a function of gate field has been interpreted in
terms of the modulation of (electron density-independent)
Altshuler-Aronov-Spivak (AAS) oscillations by the SOI,\cite{Nitta12}
emphasizing the importance of statistical averaging by the ring arrays.

In reality, however, the situation is not as simple as it appears. The
assumed ideal link of the ring to the leads is equivalent to the strong
coupling limit in terms of a connectivity parameter.\cite{Buttiker} The
implied absence of backscattering is at odds with the interpretation of the
observed oscillations in terms of AAS oscillations due to coherent
backscattering.\cite{NitConf,Nitta12} Furthermore, the experimental samples\cite%
{Bergsten,Nitta12} were not single rings in the one-dimensional quantum
limit, but a large array of connected rings, each containing several
transport channels. The tuning of the Rashba spin-orbit parameter is
associated with a strong variation in the electron density\cite{NitConf} and
therefore wave number of the interfering electrons. In the present Rapid
Communication we offer an explanation of the robustness of the observed AC
oscillations with respect to the complications summarized above.

\begin{figure}[h]
\centering
\includegraphics[trim= 0cm 4cm 0cm 6cm, clip, width=0.4\textwidth]{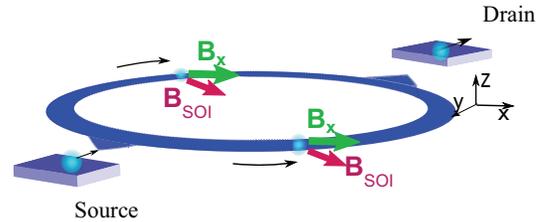}
\caption{Schematic of a quantum ring weakly coupled to source and drain contacts in the presence of SOI
effective field, $B_{SOI}$, and in-plane magnetic field, $B_{x}$.}
\label{ring}
\end{figure}

A quantitative analysis of the multi-mode ring array is very challenging and
requires large scale numerical simulations.\cite{Henri} Here we proceed from
a single single-mode quantum ring,\cite{Diego} taking backscattering into
account by assuming weak coupling to the electron leads. Its conductance can
be understood as resonant tunneling through discrete eigenstates at the
Fermi energy\cite{Buttiker} that are modulated by the SOI\ Rashba parameter.
In-plane magnetic field\cite{Nittaprivcomm} allows tuning of the
conductance oscillations without interference of the AB oscillations (see
Fig.\ \ref{ring}). We consider a modulation of the Rashba interaction
strength that is associated with an experimentally known large change in the
electron density.\cite{NitConf} Small deviations between different rings in
nanofabricated arrays can be taken into account by an ensemble averaging over
slightly different single rings. We find that this procedure leads to
an agreement with experiments that rivals previous theories.
\begin{figure}[h]
\centering
\includegraphics[trim=3.5cm 11.4cm 7.6cm 7.8cm, clip, width=0.48\textwidth] {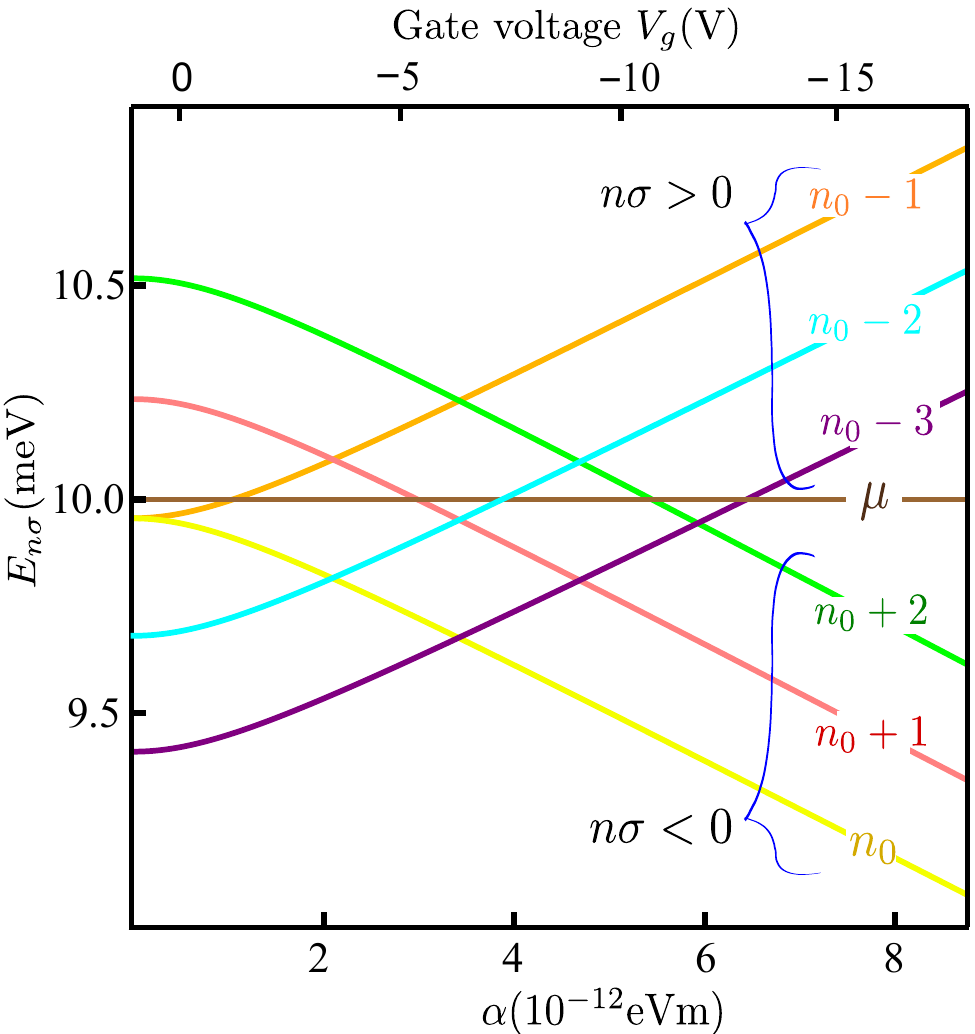}
\caption{Energies of a quantum ring with the radius $R=630$ nm close to the
Fermi energy $\protect\mu =10\,${meV} as function of the SOI strength $%
\protect\alpha $. { Energies are labeled as $n_0+i$ and $n_0 = 72$, for $n>0$, whereby, each level is Kramers degenerate with
$-n_0-i-1$ and opposite spin direction.
% Henceforth, we only label the counter clockwise moving electrons with n>0, and keep in mind the degeneracy.
}The effective
mass for InGaAs, $m=0.045\ m_{0},$ where $m_{0}$ is the electron mass. The
conductance is nonzero when $\protect\mu $ crosses an energy level.}
\label{E-al}
\end{figure}
We consider a ring with a radius of $R$, defined in the high-mobility
two-dimensional electron gas in the $x$-$y$-plane. The Rashba SOI with
the strength $\alpha $ is tunable by an external gate potential. The Hamiltonian
of an electron in the ring has the form\cite{Meijer}
\begin{align}
\hat{H}_{1D}^{(0)}& =\frac{\hbar ^{2}}{2mR^{2}}\left( -i\frac{\partial }{%
\partial \varphi }\right) ^{2}-\frac{\alpha }{R}\left( \cos \varphi \hat{%
\sigma}_{x}+\sin \varphi \hat{\sigma}_{y}\right) \left( i\frac{\partial }{%
\partial \varphi }\right)  \notag \\
& -i\frac{\alpha }{2R}\left( \cos \varphi \hat{\sigma}_{y}-\sin \varphi \hat{%
\sigma}_{x}\right) ,  \label{hla}
\end{align}%
where $m$ is the effective mass, $\varphi $ is the azimuthal angle, and $\hat{%
\sigma}_{i}$ are the Pauli matrices in the spin space. The eigenstates are
\begin{equation}
E_{n\sigma }^{(0)}=E_{R}\left[ \left( n+\frac{1}{2}\right) ^{2}+\frac{1}{4}%
+\sigma \frac{n+\frac{1}{2}}{\cos \theta }\right] ,  \label{en}
\end{equation}%
where $E_{R}=\hbar ^{2}/(2mR^{2})$, $\tan \theta =2mR\alpha /\hbar ^{2}$,
the integer $n$ is the angular momentum quantum number, and $\sigma =\pm $
denotes the spin degree of freedom.

In-plane magnetic field $B$ along the $x$-direction contributes the
Zeeman energy $H^{\prime }=E_{B}\hat{\sigma}_{x},$ where $E_{B}=g\mu _{B}B/2$%
, $\mu _{B}$ is the Bohr magneton, and $g$ is the effective g-factor. We
assume that the Zeeman energy is small compared to the (kinetic) Fermi
energy and is treated as perturbation to the zero-field Hamiltonian, $%
H_{1D}^{(0)}$.
%Figure 1 shows the spin eigenstates for the clockwise and anticlockwise moving electrons.

To leading order in $E_{B}$ the energies $E_{n\sigma }^{(0)}$ are shifted by
the in-plane field as:
\begin{align}
\Delta _{n\sigma }^{(2)}=& \frac{E_{B}^{2}}{8E_{R}}\left[ \frac{\sin
^{2}\left( 2\theta \right) }{\left( 2n\cos \theta +\sigma \right) \left(
2(n+1)\cos \theta +\sigma \right) }\right.   \notag \\
& +\left. \frac{4\sin ^{4}\frac{\theta }{2}\cos \theta }{n\left( \cos \theta
+\sigma \right) }-\frac{4\cos ^{4}\frac{\theta }{2}\cos \theta }{\left(
n+1\right) \left( \cos \theta -\sigma \right) }\right] .  \label{delta}
\end{align}%
%
%
%
%\textcolor{blue}{Since the above term should stay a perturbation to the energy, this term is only valid when $E_B^2 \ll E_a^2 n^3 \theta^2$.}

The gate voltage $V_{g}$ modifies the asymmetry of the electron confinement
potential, thereby modulating the Rashba SOI strength $\alpha $. We discuss
here first the effects of varying SOI for constant Fermi energy and
subsequently take the gate-induced density variation into account. In the absence of a magnetic field, the energy levels move
with $\alpha $ according to Eq.\ (\ref{en}). The four-fold degeneracy in the
absence of SOI $E_{n,\sigma }=E_{n,-\sigma }=E_{-n-1,-\sigma
}=E_{-n-1,\sigma }$ is broken when $\alpha \neq 0$ into two
Kramers-degenerate doublets with $E_{n,\sigma }=E_{-n-1,-\sigma }$, see Eq. (%
\ref{en}). For $\sigma n>\left( <\right) 0$ the energy increases (decreases)
with $\alpha $ as indicated in Fig. \ref{E-al}.

\begin{figure}[h]
\centering
\includegraphics[trim= 3.85cm 10.5cm 7.5cm 8.6cm, clip,width=0.47\textwidth]{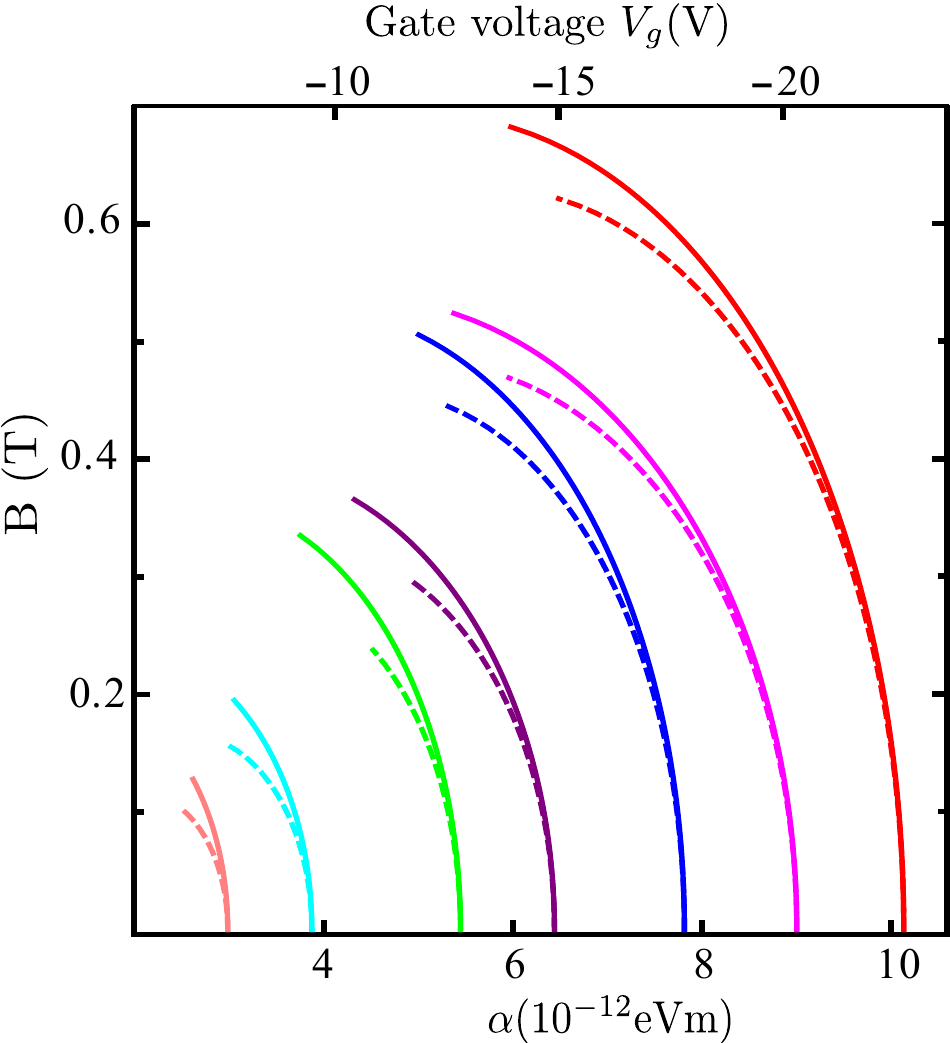}
\caption{Shift of the conductance peaks, that for zero magnetic field coincide with the crossings of the Fermi energy in Fig. 2,
 by an in-plane magnetic field as
obtained by perturbation theory. The magnetic field is seen to break Kramers
spin degeneracy. The parameters are the same as in Fig.\ \protect\ref{E-al}
and $g=-2.9$ for InGaAs.}
\label{gal}
\end{figure}

%This two-fold degeneracy is shown in the left and right hand side of Fig.
%\ref{E-al}. We assume the $n$th level is unoccupied for $E<E_F$, as shown
%in Fig.\ (\ref{fig:en}).
Resonant tunneling occurs when the energy of the highest occupied level in
the quantum ring, $E_{n_{F},\sigma }$, equals the chemical potential $\mu $
in the leads, \textit{i.e.\ } $E_{n_{F},\sigma }\left( \alpha \right) =\mu $%
, as indicated in Fig.\ \ref{E-al}. Doublets of spin-split conductance peaks
merge when $\alpha =0$, $\mu =E_{n_{F},\sigma }$, and the conductance becomes
twice as large. The in-plane magnetic field shifts the energy levels as
$\propto B^{2}$. As illustrated in Fig. \ref{gal}, the resonant
tunneling peaks at $E_{n_F,\sigma}(\alpha, B) = \mu$  are spin-split and non-parabolic. Fig. \ref{gal} 
agrees qualitatively with the experiments\cite{Nittaprivcomm} when
assuming the strong coupling limit and justifying the apparent independence
on the large $\mu $ variation with gate voltage by coherent backscattering.
In the following we suggest an alternative interpretation.

\begin{figure}[h]
\centering
\includegraphics[trim= 1.7cm 9.9cm 7.7cm 8.7cm, clip , width=0.49\textwidth] {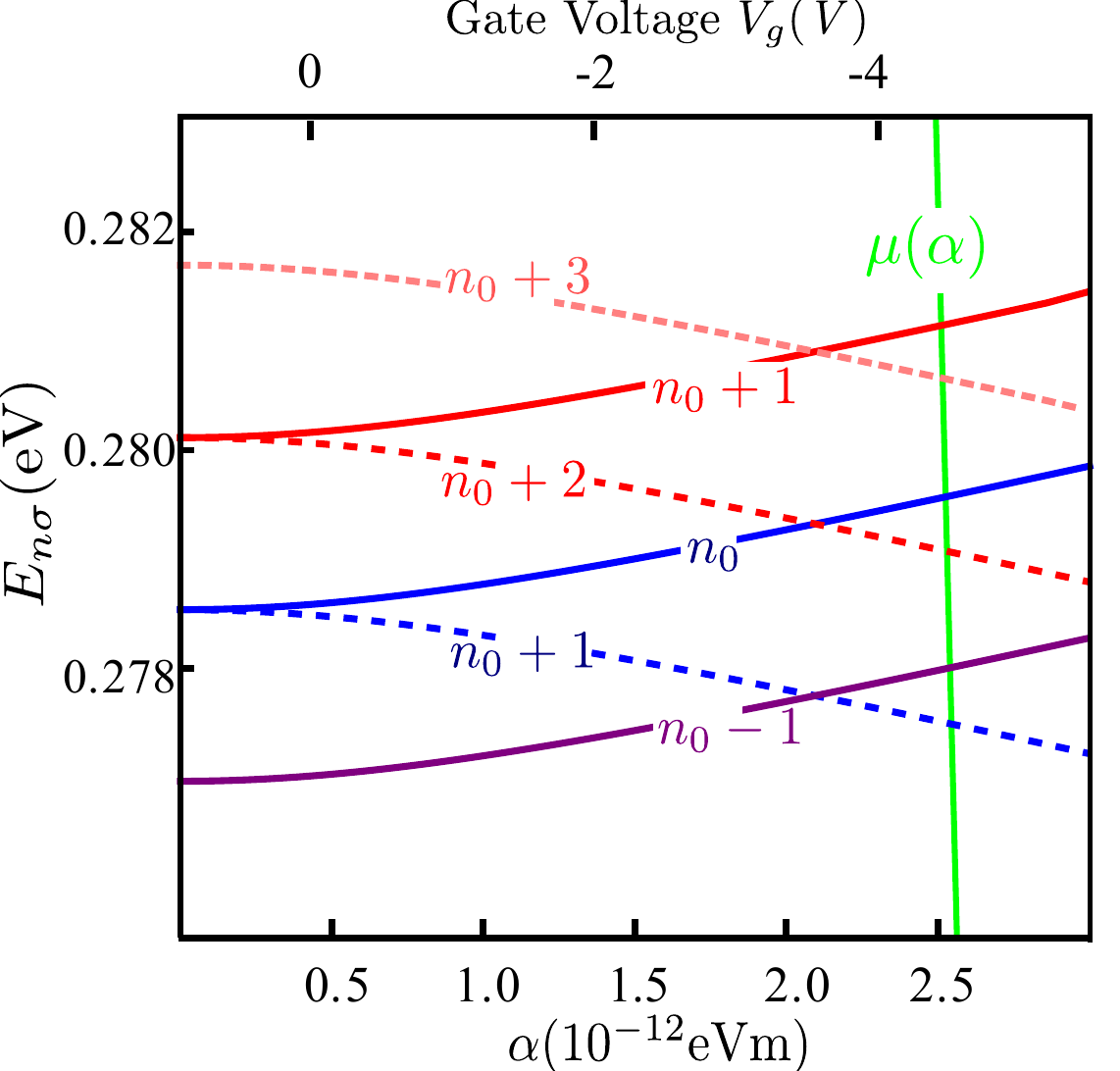}
\caption{Energy levels around the Fermi energy for $\protect\alpha %
\approx 2.4\times 10^{-12}$ eVm, compared to $\protect\mu $ that strongly depends
on the gate voltage that tunes $\protect\alpha$. {Here, $n_0=380$, dashed and solid lines represent $n\sigma<0$, and
 $n\sigma>0$, respectively. Similar to Fig. \ref{E-al}, we labeled the energies for $n>0$, keeping in mind that the energy levels are 
twofold degenerate.}}
\label{Eal}
\end{figure}

\begin{figure}[h]
\centering
\includegraphics[trim=3.6cm 10.6cm 6cm 5.4cm, clip,width=0.49\textwidth]{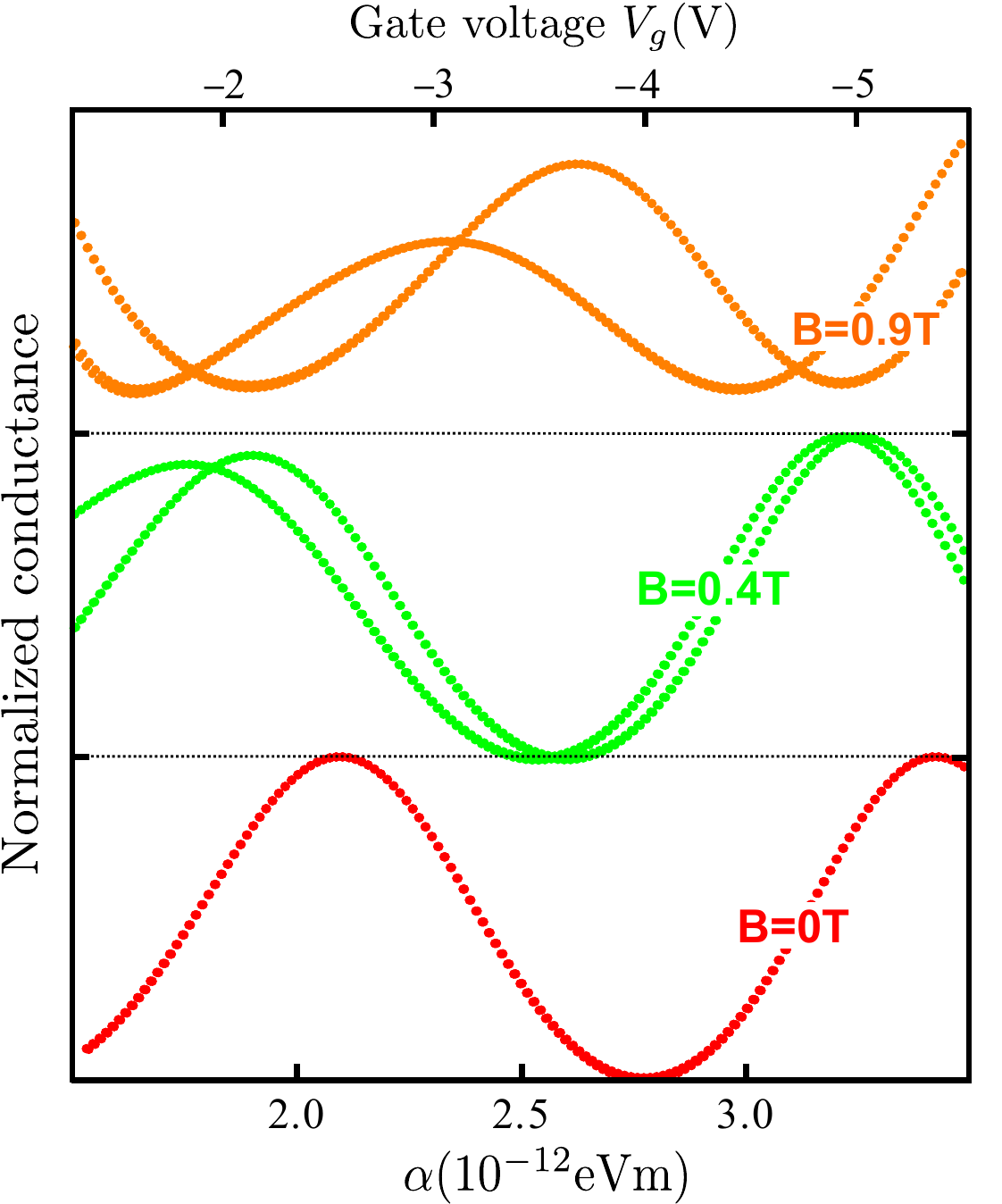}
\caption{The conductance oscillations of an ensemble of rings
with  energy levels broadened by a Gaussian with $\Gamma =0.003$ peV/m as
a function of an in-plane magnetic field.
% Results have been normalized as $%
%\Delta G_{\protect\alpha \neq 0}/\Delta G_{\protect\alpha =0}$, where $%
%\protect\alpha $ is the Rashba constant. 
{The dashed lines are guides to the eye,
to compare the oscillation amplitudes while varying the magnetic field. All amplitudes are scaled with
those at $B=0$T that display a modulation of $(G_{max}-G_{min})/G_{max}$ = 50$\%$.
% the amplitude of the oscillation in
%the absence of the magnetic field. This amplitude is obtained from $G_{max}-G_{min}$, where $G_{max(min)}$, the maximum and minimum of conductance
%in the absence of the magnetic field, are
% $G_{max}=300.086 G_0 T$, and $Gmin=145.232 G_0 T$, while $G_0 = e^2 /\hbar$ and T is the
%tunneling amplitude to and from the ring. 
}
 The in-plane magnetic field splits
Kramers degenerate spin states that evolve differently with gate voltage.}%
\label{GBn}
\end{figure}

According to the experiments,\cite{Bergsten,Nitta12} $\alpha $ depends on the
gate voltage as $\alpha \left[ 10^{-12}\,\mathrm{eVm}\right]
=0.424-0.47\times V_{G}\left[ 10^{-12}\,\mathrm{V}\right] $ and on the electron
density as $\alpha \left[ 10^{-12}\mathrm{eVm}\right] =7.81-3.32\times N_{s}%
\left[ 10^{12}\mathrm{cm}^{-2}\right] .$ In Fig. \ref{Eal} we plot the ring
energies as a function of $\alpha $ including the chemical potential $\mu \ $%
that varies much faster with $\alpha $ than the single particle energies,
leading to conductance peaks that as a function of gate voltage are very
closely spaced. In ring arrays\cite{Bergsten,Nitta12} we do not expect to
resolve such narrow resonances due to disorder, multi-mode contributions and
ring size fluctuations. We can model the latter by averaging over an
ensemble of rings with a Gaussian distribution of resonant energies or
conductance peak positions with a phenomenological broadening parameter $%
\Gamma $.{Fig. \ref{GBn} illustrates the result of the averaging procedure
in the form of the normalized conductance modulations.}\cite{note}  While the resonant tunneling peaks are smeared
out, slow (AC) oscillation as a function of $\alpha $ reappears, which
represents the beating of the level spacings induced by the SOI,
in qualitative agreement with experiments.

\begin{figure}[h]
\centering
\includegraphics[trim= 5.3cm 1.6cm 6.3cm 17.8cm, clip, width=0.49\textwidth]{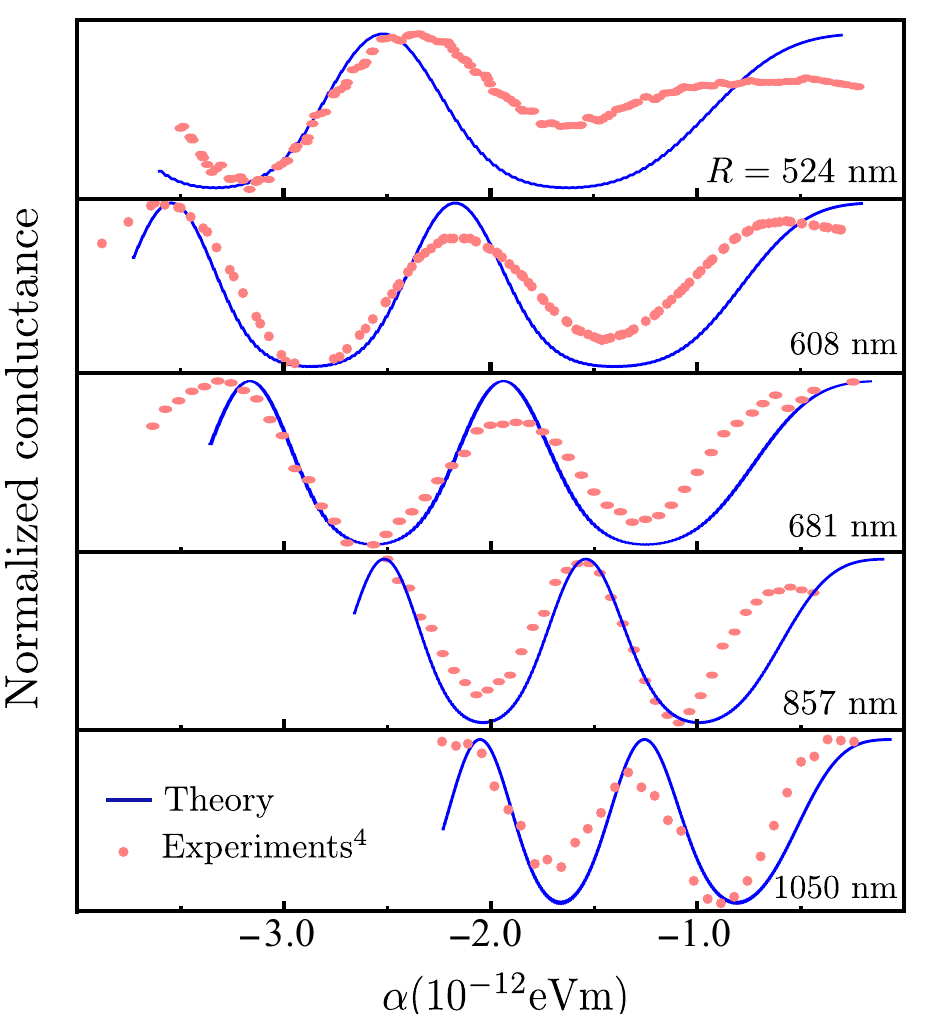}
\caption{The conductance G of a array of rings modeled as an ensemble
energy levels as a function of $\alpha$ broadened by a Gaussian
for various nominal radii R. The broadening parameters are $\Gamma =0.005,\, 0.0035,\, 0.003,\, 0.002$ and $0.001$ peV/m, for
$R= 524;\,608;\, 681;\, 857$ and $1050$ nm, respectively. All amplitudes are scaled to a panel height corresponding to $(G_{max} - G_{min} )/G_{max} = 50 \%$.
We
use the experimentally determined relations between
Rashba constant and electron density as before. 
We compare our calculations (lines) with the experimental results
(points) from Ref. \protect\onlinecite{Nitta12} (see also Ref. \protect\onlinecite{note}).}
\label{ACr}
\end{figure}
The experiments of AC oscillations in arrays with different ring radii\cite%
{Nitta12} are compared in Fig. \ref{ACr} with our results and that of a
single open ring.

In Fig.\ \ref{GBn}, we also illustrate the effect of in-plane magnetic field
on the ensemble of Rashba rings. The magnetic field shifts the phase of the
oscillation to lower values of the gate voltage or larger $\alpha $ and thus
suppresses the amplitude of the conductions oscillations increasingly for
lower values of the gate voltage. %\textbf{GB}: \textit{I do not understand
%the }$\bar{G}_{0}\ $\textit{here. The amplitude of the conductance
%modulations should be only a fraction of the total conductance. I also do
%not see if and how the amplitudes are suppressed. This could be improved by
%plotting horizontal lines in the figure. }
These features agree again qualitatively with those observed experimentally
by Nitta \textit{et. al}\cite{Nittaprivcomm}, although our theory appears to
overestimate phase shifts at large negative gate voltages. The magnetic
field splits the Kramers degeneracy, thereby leading to two sets of
superimposed oscillations that might be experimentally resolved in the
form of different Fourier components.

Previous theories\cite{Diego,Meijer} treat ideally open single rings, while
we consider the weak coupling limit. Both extremes are likely not met in
experiments. The intermediate regime can be modeled in terms of a
connectivity parameter.\cite{Buttiker} An increased coupling causes a
Lorentzian smearing of the conductance peaks, which is likely to effectively
enhance the phenomenological broadening of the ensemble average and cannot
be resolved in the experiments. The presence of several occupied modes in
the rings also contributed to the average, since each radial node can be
approximated as a ring with a slightly different radius. We therefore
believe that our results are robust with respect to deviations from our
Hamiltonian and these deviations can be captured by the phenomenological broadening parameter
$\Gamma .$

In conclusion, we investigated the conductance of single rings and an
ensemble of them as a function of the Rashba spin-orbit interaction in the
limit of weak coupling to the leads. We considered both constant and gate
voltage-dependent density of electrons. Both situations can in principle be
realized experimentally by two independent (top and bottom) gate voltages.
We compare results with experiments on ring arrays in which a single gate
changes both the SOI $\alpha $ as well as the electron density. We found
that, in agreement with experiments, the ensemble averaged conductance
oscillates as a function of $\alpha $. The oscillations undergo a phase
shift under an in-plane magnetic field,  { and the period varies with the ring diameter}, as
observed. We conclude that experiments observe SOI-induced interference
effects that are more complicated than the original Aharonov-Casher model
but are robust with respect to- the model assumptions.

We would like to thank Prof. J. Nitta for sharing his unpublished data. This
work was supported by FOM (Stichting voor Fundamenteel Onderzoek der
Materie), EU-ICT-7 \textquotedblleft {MACALO}\textquotedblright {,}\ the
ICC-IMR, and DFG Priority Programme 1538 \textquotedblleft {Spin-Caloric
Transport}\textquotedblright\ (GO 944/4).

%\bibliography


\begin{thebibliography}{99}
\bibitem{note}
Experiments report resistances rather than conductances. The resistance oscillations $\delta R\ll R$ are
measured on top of a large background resistance $R$. 
Thus, experiments are represented by $\delta R (\alpha \neq 0) / \delta R (\alpha =0) = \delta G( \alpha \neq 0) / \delta G (\alpha =0)$,
 which we may compare directly with our calculated normalized conductance oscillations.
\end{thebibliography}
\end{document}